# Improvement of HRS Variability in OxRRAM by Tailored Metallic Liner

N. Guillaume[1]*, M. Azzaz[1,3], S. Blonkowski[1,3], E. Jalaguier[1], P. Gonon[2], C. Vallée[2], T. Blomberg[4], M. Tuominen[4], H. Sprey[5] S. Bernasconi[1], C. Charpin-Nicolle[1], E. Nowak[1]

[1] CEA LETI, MINATEC Campus
Grenoble 38000, France
E-mail: nicolas.guillaume@cea.fr

[2] LTM, MINATEC Campus
Grenoble 38000, France

[3] STMicroelectronics

Crolles 38920, France
[4] ASM Microchemistry
[5] ASM Belgium

**Abstract**

"In this work, we propose a novel integration in order to significantly reduce the High Resistance State variability and to improve thermal stability in Oxide-based Resistive Random Access Memory (OxRRAM) devices. A novel device featuring a metallic liner, acting as a parallel resistance, is presented. To assess the effect of this solution, we compare the results with a standard OxRRAM cell structure. A very good stability of the resistive states, both in endurance and temperature, is highlighted and explained thanks to a conductive filament based model."

## 1. Introduction

In the upcoming challenges in order to shrink memory size while improving their characteristics [1], resistive memories are serious candidates. Resistive memories are characterized by two main resistive states, HRS and LRS (respectively High and Low Resistance State) and are declined into many categories. Among them, we focused on MIMs (Metal-Insulator-Metal) structured OxRRAM (Oxide Resistive Random Access Memory). OxRRAMs, based on the formation (set) and rupture (reset) of conductive filament, already proved their potential in terms of switching speed, consumption, endurance, retention and scaling capability [2] but still need to be improved in terms of variability, mainly for HRS [3]. This variability can be quantitatively explained through the tunneling or field assisted tunneling current between two parts of the filament when it is disrupted [4].

In this paper, we propose to investigate the impact of a metallic liner on the variability of the HRS and the thermal stability of the LRS. The purpose is adding a parallel resistance and transferring a part of the electrical current and of the thermal stress on the liner during switching operations. This technique has already been used in Phase Change Memory to lower the drift of the HRS [5]. Fig. 1 represents two studied memory stacks, one classic OxRRAM stack (Device I) and the one integrating the metallic liner (Device II). Thanks to electrical characterizations, we provide the demonstration of the reliability of our solution. Moreover, a physical model is proposed to explain our results and the advantage of the proposed solution.

## 2. Methods and integration

The technology used for the memory stacks is described as follow. The memory stack is composed of PVD TiN as bottom electrode, 10 nm thick ALD $HfO_2$ as oxide layer, then 10 nm of PVD Ti and 50 nm of PVD TiN as top electrode. The metallic liner is grown by ALD using a Pulsar™ reactor from ASM and consists of a layer of TiWN [6]. The schematic of the filament configuration is represented in Fig. 2 alongside the description of the model parameters. All the devices tested in this paper were coupled with a transistor (1T1R devices) and were polarized in order to get a compliance current of 300μA for all devices.

## 3. Results and discussion

*Pulsed endurance and retention measurements*

Endurance measurements are represented in Fig. 3 after the initial forming operation leading to the LRS. Resistance measurements, at 0.1V, in high and low states with their respective deviation are listed in Table I. Device I shows a larger memory window, but the HRS is unstable (Standard Deviation of the HRS is roughly 45%). Conversely, device II has a narrow memory window (about 1 decade), which is acceptable, and dispersion of the HRS is almost negligible compared to device II, since its standard deviation is lowered to 4%.

Fig. 4 shows the retention measurements that were performed on devices set in LRS at the temperature of 250°C. As expected, device II allows a good stability of the LRS thanks to the thermal stress release through the metallic liner whereas device I shows an important variability on several devices and start to drift at $10^5$s.

*Discussion and model*

The basis of the model is to compute the time evolution of a state variable P that accounts for the filling of a constriction, whose maximum area is Sc in the OxRRAM conductive filament (Figure 2 left). P=1 corresponds to LRS, where the current is a linear function of the voltage [7]. When P=0, the constriction is open and corresponds to the HRS so the current is due to tunneling in the gap whose length is d. The filament in the HRS state has been schematized together with the liner (right part of Figure 2). Using the same set of parameters for $HfO_2$ based similar OxRRAM from [7] we calculated the quasi static current voltage characteristic. For device II, a parallel resistance of 12.9kΩ was self consistently coupled to the same model with the same parameters.

Fitted results of the model are represented alongside quasi-static measurements performed on devices I and II in Figure 5. For device I the HRS current is due to tunneling in the gap and is strongly nonlinear with a typical order of magnitude value of $R_{HRS}\sim 50k\Omega$. For device II, in the LRS state, the current mainly flows in the filament similarly to the device I, while in the HRS state, the current mainly flows in the parallel liner resistance of $12.9k\Omega$, which is lower than $R_{HRS}$. This clearly explains the narrowing of the memory window for device II. The current statistical fluctuations are much larger for tunneling [8] compared to the fluctuations of the On state and the liner which correspond to the current in a continuous conductor. For device II because the fluctuations come from the continuous liner, the Off states HRS fluctuations are very small. The choice of the parallel resistance value is important. This value has to be higher than the LRS one and lower than the HRS one. It must be larger than the quantum point resistance, which corresponds to the highest resistance value reachable in the LRS. For a larger parallel resistance, the HRS fluctuations will contribute more significantly.

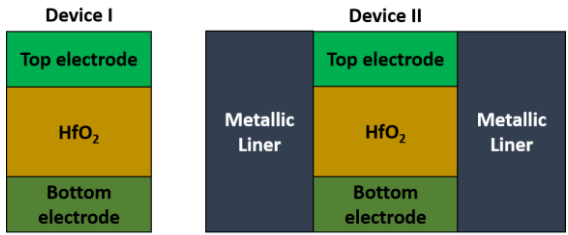

Fig. 1 Schematic representation of the two devices tested in this work, a classic memory stack (left) and a memory stack with a metallic liner (right)

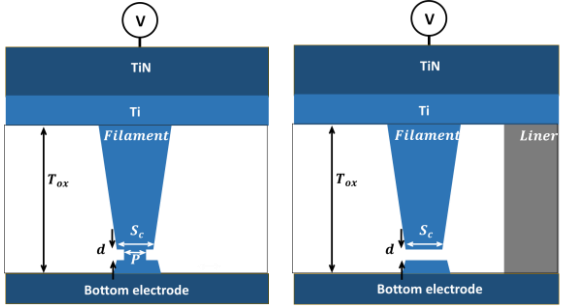

Fig. 2. Schematic view of the filament model. Tox stands for the oxide thickness. Standard device I representation (left after [6]). Device II in the HRS state representation (right), after [7].

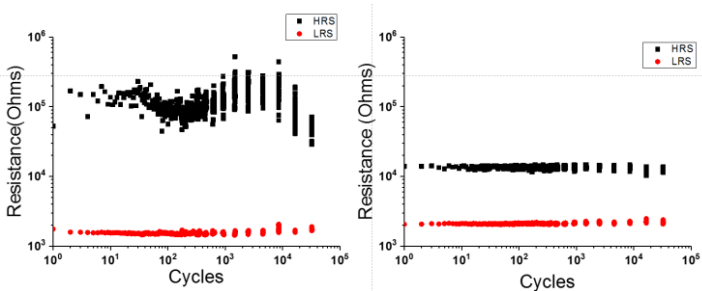

Fig. 3. Endurance results in pulse mode on device I (left) and device II (right) for 32k cycles.

Table I Endurance results in pulsed mode

| Devices | Mean value of LRS (kOhms) | Mean value of HRS (kOhms) | Standard Deviation (SD) LRS | Standard Deviation (SD) HRS |
|---|---|---|---|---|
| Device I | 1.6 | 120 | 6% | 45% |
| Device II | 2.1 | 14 | 3% | 4% |

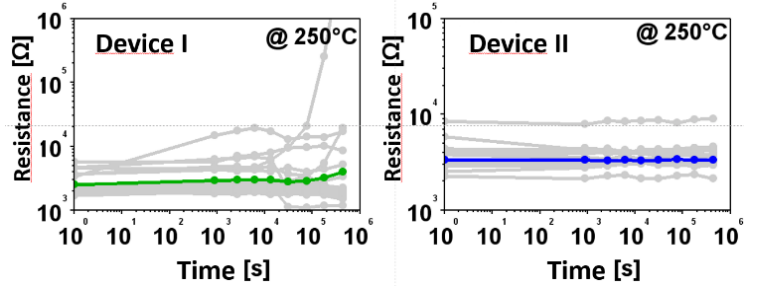

Fig. 4. Evolution of memory's resistance as a function of time at 250°C for devices set in LRS and for both device I (left) and device II (right). Blue and green lines represent the median and grey lines represent all the devices tested

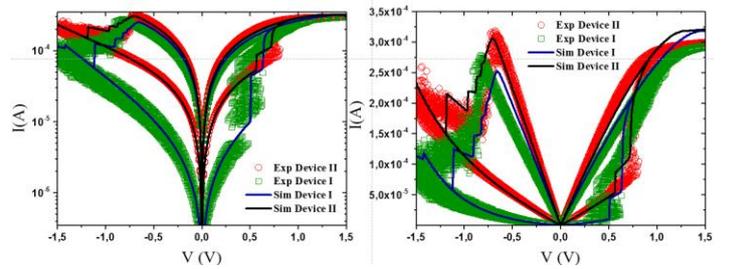

Fig. 5. Experimental data from reset/set cycling in quasi-static for both device II (red circle) and device I (green square) fitted by the filamentary model for both cases (respectively blue and black lines). Curves are presented either in log scale (left) or in linear scale (right).

## 4. Conclusions

In this paper, thanks to the addition of a metallic liner, we provided an experimental evidence on 1T1R devices of HRS fluctuation drastic reduction that give rise also to a memory window narrowing. The metallic liner implementation allowed the memory to cycle with a very good stability of its both states HRS and LRS (SD=4%) with also a very good stability in retention. Finally, it has been found that the parallel resistance needs to be close to the quantum resistance.

## 5. Acknowledgements

This work has been supported by the European ESCEL JU within the PANACHE and WAKeMeUP projects.